\documentclass[notitlepage,czech,english,12pt,a4paper]{article}
\usepackage{setspace}
\usepackage{epsfig}

\usepackage{feynmf}
\usepackage{slashed}

\usepackage{babel}
\usepackage[T1]{fontenc}
\usepackage[cp1250]{inputenc}

\newcommand{\be}{\begin{equation}}
\newcommand{\ee}{\end{equation}}
\newcommand{\bea}{\begin{eqnarray}}
\newcommand{\eea}{\end{eqnarray}}

\newcommand{\ep}{i\varepsilon}
\newcommand{\nn}{\nonumber}

\begin{document}
\begin{center}
\vspace*{1.4cm} {\LARGE \bf Towards the solution of Schwinger-Dyson equations in Minkowski space}\\[2cm]
{\LARGE Vladimír Šauli}\\[1cm]
{\large Faculty of Mathematics and Physics}\\
{\large Charles University, Prague}\\[0.2cm]
{\large{\center and}}\\[0.3cm]
{\large Nuclear Physics Institute}\\
{\large Czech Academy of Sciences, Řež near Prague}\\[1.5cm]
{\large Branch f-9 -- Subnuclear physics}\\[3cm]
{\large Abstract of author's PhD Thesis 'Schwinger-Dyson approach\\[0.5cm]
 to field models with strong couplings'}\\[1.5cm]
{\large April  2005}
\end{center}

\newpage\thispagestyle{empty}

\pagenumbering{arabic}\setcounter{page}{1}


\section{Introduction}

This PhD thesis is devoted to studies of  quantum field models
with strong coupling. The {\em Schwinger-Dyson equations} (SDEs)
in momentum representation are solved in Minkowski space. The
gauge invariant non-perturbative regularization scheme is
constructed  to treat correctly  non-asymptotically free models
and propagators of the strong-coupling QED in 3+1 dimension are
calculated. As for QCD, our method in its present form is not able
to deal with the Faddeev-Popov ghosts. However, the analyticity of
the gluon propagator is exploited to continue some recent lattice
data to the timelike momentum axis.

In this introduction our motivations, broad contexts and structure of
the thesis are briefly summarized.

These days the experimentally accessible particle physics is being
described in the framework of the (effective) quantum field theory
(QFT) by the Standard model of the strong and electroweak
interactions. In the perturbation theory (PT)  the S-matrix elements
are generated from n-point {\em Green's functions} (GFs) which are
calculated and regularized/renormalized order by order according to
the standard rules. Order by order PT can be naturally generated from
expansion of the SDEs: an infinite tower of coupled equations
connecting successively higher and higher points GFs.

Going beyond the PT requires development of elaborated approaches
and sophisticated tools, especially in the strong coupling regime.
Number of such approaches has been pursued: lattice theories,
Feynman-Schwinger representation, non-perturbative treatments of
SDEs, bag models \dots. These days neither of them provides such a
clear and unambiguous understanding of the physics as we are
accustomed to in the perturbative regime.

In this work we deal with the non-perturbative solutions of the
SDEs. If one could solve the full infinite set of SDEs, the
complete solution of the QFT will be available. In reality one has
to truncate this set of equations: the main weakness of SDEs
phenomenology is the necessity to employ some Ansatze for higher
Green functions.   The reliability of these approximations can be
estimated by comparison with the known PT result in the regime of
a soft coupling constant and by comparison with the lattice data
(and/or results of alternative approaches) where available. The
recent results (see e.g.\ reviews
\cite{ROBERTS,ALKSMEK2001,MARROB2003,FISALK2003} and references
therein) provide some encouragement: they suggest that SDEs  are
viable tool for obtaining (eventually) the truly non-perturbative
answers for plethora of fundamental questions: {\it dynamical
chiral symmetry breaking}, {\it confinement} of colored objects in
QCD, the high temperature superconductivity in condensed matter
(modeled \cite{MAPA2003} by $QED_{2+1}$). In the QCD sector there
is also a decent agreement with some experimental data: for meson
decay constants like $F_{\pi}$, evolution of quark masses to their
phenomenologically known (constituent) values, bound states
properties obtained from Bethe-Salpeter or Faddeev equations (the
part of SDEs).  The recent solutions of SDEs
\cite{FISALK2003,BLOCH2003} nicely agree with the lattice data
\cite{BHLPW2004,BOBOLEWIZA2001,BCLM2002} (the explicit comparison
is made in \cite{FISALK2003,BLOCH2003,FIES2004}). Thus,
encouraging connections between results of SDEs studies and
fundamental theory (QCD) and/or various phenomenological
approaches (chiral perturbation theory, constituent quark model,
vector meson dominance \dots) are emerging.

Most of these results were obtained by solution of the SDEs in
Euclidean space. One of the main goals of this work is to develop an
alternative approach allowing to solve them directly in Minkowski
space. In this method (generalized) {\em spectral decompositions} of
the GFs are employed, based on their analytical properties. The main
merit of this approach is possibility to get solutions both for
spacelike and timelike momenta. Techniques of solving SDEs directly
in Minkowski space are much less developed than the corresponding
Euclidean ones. Therefore, most of this thesis deals with developing
and testing such solutions on some simple QFT models.

For large class of {\em non-asymptotically free} (NAF) theories
running couplings -- calculated from PT -- diverge at some finite,
but usually large, spacelike scale.  This is why  we pay special
attention to behavior of our NAF solutions in the region of large
momenta.

One should mention a possibility of getting solutions of the SDEs in
Minkowski space with the help of analytical continuation of the
Euclidean ones. In simple cases this can be successful: we have
recently done this for the gluon form factor and we present the
results in this thesis in Section devoted to the QCD. We have found
the non-positive absorptive part contribution to the gluon propagator
in the Landau gauge, in agreement with some recent analyzes.

The thesis starts with an introduction of essential ideas of the
integral representation in QFT, in particular of the Perturbation
Theory Integral Representation (PTIR), and with discussion of the
renormalization in the framework of non-perturbative SDEs. In
Chapters~2 and 3 of the thesis some well-known general results and
useful textbooks formulas are collected and basics of the
formalism employed are discussed. The derivation of SDEs is
reminded for the well-known example of QED. Chapter~3 deals with
integral representation for various GFs. The spectral
representation of one-particle GF is derived.  Then the
generalization to higher point Greens function is presented with
some examples. One of these examples -- the derivation of the
dispersion relation for the sunset diagram -- is original work of
the author.

Then, the technique based on the integral representation of  GFs
is used to solve SDEs for several quantum field models. Chapter~4
is devoted to the simple scalar model. The SDEs are briefly
rederived and then the momentum space SDE is converted into the
{\it Unitary Equation} (UE) for propagator spectral density. The
UE is shown to be a real equation, in which  the singularities
accompanying usual Minkowski space calculations are avoided.  The
relativistic bound-state problem as an intrinsically
non-perturbative phenomenon is treated in Chapter 5.

Large part of the thesis is devoted to  solutions of analogous
problems in more complicated theories. The systems of the SDEs for
QED is considered in two approximation, one of which, the  so-called
ladder approximations, is widely used as a pedagogical introduction
to the subject. The second approximation employs the Ball-Chiu
vertex, consistent with the Ward-Takahashi identity. These studies
are based on papers \cite{SAULIJHEP,SAULIRUN}. Further insights and
improvements of rather technical character can be found in  recent
author's paper \cite{SAULI2005}, where some obstacles due to singular
integral kernels are avoided.

The  Yukawa theory, i.e., the quantum field theory describing the
interaction of spinless boson with Dirac fermion is considered in
rather simple bare-vertex approximation in Chapter 7  of the thesis.
A generalized spectral Ansatz is employed to obtain the spectral
function from recent lattice data (and recent solution of SDEs for
gluon propagator in Euclidean space) in Chapter 8.

In what follows we briefly summarize the main results of the thesis,
including also some basic information on current status of particular
problems.

\section{Scalar toy-model}

We solve the SDEs for scalar $\Phi^3$ and $\Phi_i^2\Phi_j$
theories, the second model is referred to as the (generalized
massive) Wick-Cutkosky model (WCM).  The solution is performed
with the help of the spectral technique directly in Minkowski
space. The first aim of this exercise is to test the employed
method of solution of integral equations (by iterations) and its
actual implementation (numerical accuracy) by varying numerical
procedure and by comparing to results obtained in Euclidean space.
The second aim is to obtain the dressed propagators to be used in
the bound-state studies. The SDEs have been solved in the bare
vertex approximation and also with non-trivial improvement of the
interaction vertex. We discuss {\it multiplicative
renormalization} in its non-perturbative context. Our method
provides the solutions only for coupling constants smaller than
certain critical value. It is explained that the method fails when
the field renormalization constant approaches zero.

The (metastable) scalar models  often serve as an useful
methodological tool. The main reason is, of course, their inherent
simplicity: absence of spin degrees of freedom leads to lowest
possible number of independent amplitudes and the $\Phi^3$-like
models involve also simplest possible vertices. The $\Phi^3$
theory has already been employed as a suitable playground for
studies of various phenomena
\cite{COR1995,COR1997,DEM1997,BROKRE2001}, including
non-perturbative asymptotic freedom  and non-perturbative
renormalization.

Dealing with massive theories we will use the generic spectral
decomposition of the renormalized propagator of the stable and
unconfined particle:
\be G(p^2)= \frac{r}{p^2-m^2+i\epsilon}+
\int\limits_{\alpha_{th}}^\infty d\alpha\,
\frac{\sigma(\alpha)}{p^2-\alpha+\ep} \, , \ee
where - as  will be confirmed by actual evaluation - the spectral
function $\sigma(\alpha)$ is positive, regular and  is spread
smoothly from zero at the threshold  $\alpha_{th}=4m^2$.

Putting the spectral decomposition of the propagators and the
expression for the vertex function into the SDE allows one to derive
a real integral equation for the weight function $\sigma(\alpha )$.
This equation can be solved numerically by iterations. Since  all
momentum integrations are performed analytically, there is no
numerical uncertainty following from the renormalization, which is
usually not the case in the Euclidean formalism. In the thesis the
renormalization procedure is performed analytically with the help of
direct subtraction in momentum space. The
super-renor\-ma\-li\-za\-bility  makes our models particularly
suitable for model studies. It implies the finiteness of the field
renormalization constant $Z$, which therefore  need not be considered
at all. Nevertheless, for the $\Phi^3$ theory  the field
renormalization is not fully omitted, but with the help of an
appropriate choice of the finite constant $Z$ we choose the
renormalization scheme. Minkowski results obtained in the bare vertex
approximation and with trivial field renormalization $Z=1$ are
compared with the appropriate Euclidean solutions.

In the Section devoted to  $\Phi^3  $ theory
it is shown how to properly renormalize the SDEs. The key
point is the requirement of multiplicative renormalization, then by
the construction the S-matrix puzzled from the GFs is invariant under
the choice of renormalization schemes in which these GFs have been
calculated.

The Lagrangian density for this model  reads
\be \label{lagr}
{\cal L}(x)=\frac{1}{2}\partial_{\mu}\phi_0(x)\partial^{\mu}\phi_0(x)-
\frac{1}{2}m_0^{2}\phi_0^{2}(x)-g_0\phi_0^{3}(x) \, ,
\ee
where the subscript $0$ indicates the unrenormalized  quantities.

The propagator SDE in the momentum space reads:
\begin{fmffile}{pi0}
\bea  \label{UDSE}
G_0^{-1}(p^2)&=&p^2-m_0^2-\Pi_0(p^2) \, ,
\nn \\
\Pi_0(p^2)&=&
\parbox{3.0\unitlength}{%
\begin{fmfgraph}(3.0,2.5)
\fmfpen{thick}
\fmfleft{i}
\fmfright{o}
\fmf{plain}{i,v1}
\fmfblob{0.1w}{v1}
\fmf{plain}{v2,o}
\fmf{plain,left,tension=0.25}{v1,v2}
\fmf{plain,right,tension=0.25}{v1,v2}
\end{fmfgraph}}
\quad \quad = \nonumber\\
&& i3g_0^2\, \int\frac{d^dq}{(2\pi)^d}\,
\Gamma_0^{[3]}(p,q-p,-q)G_0(q-p)G_0(q) \, , \eea
\end{fmffile}
where the arguments of $\Gamma_0^{[3]}(p_1,p_2,p_3)$ are the incoming momenta
(and $p_1+p_2+p_3=0$).

\begin{figure}
\begin{fmffile}{pi3}
\bea
\Gamma^{[3]}(p,l-p,-l)&=&\,\,\,\,\,\,\,
\parbox{2.0\unitlength}{%
\begin{fmfgraph}(2.0,1.5)
\fmfpen{thick}
\fmfleft{i}
\fmfrightn{o}{2}
\fmfblob{0.1w}{v}
\fmf{plain}{i,v}
\fmf{plain}{v,o1}
\fmf{plain}{v,o2}
\end{fmfgraph}}
\\
& = &\,\,\,
\parbox{2.0\unitlength}{%
\begin{fmfgraph}(2.0,1.5)
\fmfpen{thick}
\fmfleft{i}
\fmfrightn{o}{2}
\fmf{plain}{i,v}
\fmf{plain}{v,o1}
\fmf{plain}{v,o2}
\end{fmfgraph}}
\, \, \, + \, \, \,
\parbox{2.0\unitlength}{%
\begin{fmfgraph}(2.0,1.5)
\fmfpen{thick}
\fmfleft{i}
\fmfrightn{o}{2}
\fmf{plain}{i,v}
\fmf{plain,left,tension=0.25}{v,v2}
\fmf{plain,right,tension=0.25}{v,v2}
\fmfv{d.sh=square,d.f=hatched,d.si=0.25w,l=$\Gamma_4$}{v2}
\fmf{plain}{v2,o2}
\fmf{plain}{v2,o1}
\end{fmfgraph}}
\, \, \, + \, \, \,
\parbox{2.0\unitlength}{%
\begin{fmfgraph}(2.0,1.5)
\fmfpen{thick}
\fmfleft{i}
\fmfrightn{o}{2}
\fmf{plain}{i,v}
\fmf{plain}{v1,o1}
\fmf{plain}{v2,o2}
\fmfblob{0.1w}{v1}
\fmfblob{0.1w}{v2}
\fmf{plain,tension=.3}{v1,v}
\fmf{plain,tension=.2}{v1,v2}
\fmf{plain,tension=.3}{v2,v}
\end{fmfgraph}}
\nonumber \\
&\simeq&
\parbox{2.0\unitlength}{%
\begin{fmfgraph}(2.0,1.5)
\fmfpen{thick}
\fmfleft{i}
\fmfrightn{o}{2}
\fmf{plain}{i,v}
\fmf{plain}{v,o1}
\fmf{plain}{v,o2}
\end{fmfgraph}}
\, \, \, + \, \, \,
\parbox{2.0\unitlength}{%
\begin{fmfgraph}(2.0,1.5)
\fmfpen{thick}
\fmfleft{i}
\fmfrightn{o}{2}
\fmf{plain}{i,v}
\fmf{plain}{v1,o1}
\fmf{plain}{v2,o2}
\fmf{plain,tension=.3}{v1,v}
\fmf{plain,tension=.2}{v1,v2}
\fmf{plain,tension=.3}{v2,v}
\end{fmfgraph}}
\nonumber
\end{eqnarray}
\end{fmffile}
\caption{Diagrammatical representation of the SDE for the vertex
function. As in previous figure all internal propagators are dressed.
Blobs represent the full vertex, the box stands for $\Gamma^{[4]}$.}
\end{figure}

In the thesis the  system of SDEs is closed already at the level of
the equation for the proper vertex. Replacing the full vertex by its
tree approximation defines the {\it bare vertex approximation} (BVA),
in our {\it dressed vertex approximation} (DVA) the next term of the
skeleton expansion of the SDE for the vertex is also included. The
equation for the propagator is solved in the BVA and DVA for two
renormalization schemes.  The on-mass-shell subtraction  is employed
to  renormalize the self-energy, for an extension to the off-shell
renormalization scheme see \cite{SAULI2005}.

After the renormalization  some simple algebra allows us to convert
the momentum space SDE (\ref{UDSE}) into two equations for $\sigma$
and the absorptive part of self-energy $Im\ \Pi= \pi \rho$. The
latter is obtained from the dispersion relation for renormalized
self-energy, which e.g.\ for the Minimal momentum subtraction (MMS)
renormalization  scheme reads:
\be
\Pi_R(p^2)= \int\limits_{4m^2}^{\infty}d\alpha \,
\frac{\rho(\alpha)(p^2-m^2)}{(p^{2}-\alpha+i\epsilon)(\alpha-m^2)}
\quad ,
\label{vrku1}
\ee
where $m$ is the ``physical'' (pole) mass
of the scalar.

Having solved the system of equations for $\sigma$ and $\rho$ the GFs
are  calculated through their integral representation. This is the
essence of the spectral approach to SDEs in Minkowski space.

For the MMS RS the comparison with more usual Euclidean solution was
made. Up to rather small (numerical) deviation the results for
propagators agree in the spacelike regime, where solutions are
available for both approaches (Fig.~1). Also the comparison to the
usual perturbation theory was made and the critical value of the
coupling constant was determined.

\begin{figure}[t]
\centerline{\epsfig{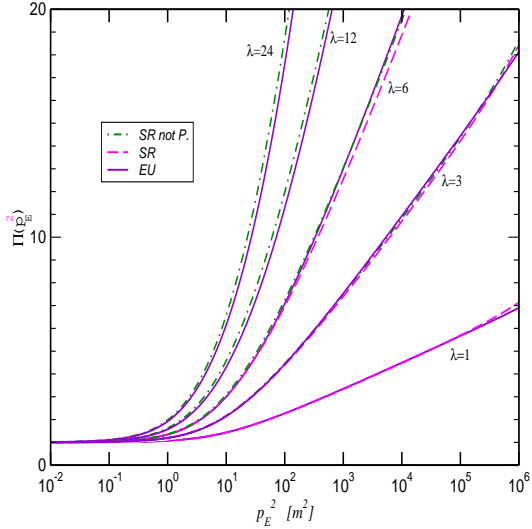}}
\caption[99] {The function $\Pi(p_E^2)$ calculated in the spectral
(dashed line) and Euclidean formalism (solid line). Comparison is
made for various couplings $\lambda=\{1,3,6\}$, for the strongest
case $\lambda=12, 24$ the spectral solution has not been found. The
dashed line shows the solution of the SDEs in which the principal
value integrals were omitted. \label{fig6.1}}
\end{figure}

\subsection{The relativistic bound-states}

\begin{figure}[t]
\centerline{  \mbox{\psfig{figure=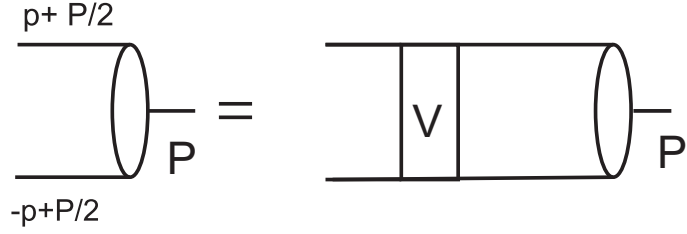,height=3.0truecm,angle=0}} }
\caption[99]{ \label{figBSE}Diagrammatic representation of the BSE for
the bound state vertex function.}
\end{figure}

In  quantum field theory the two-body bound state is described by the
bound state vertex function or, equivalently, by Bethe-Salpeter (BS)
amplitude, both of  them are solutions of the corresponding (see
Fig.~\ref{figBSE}) covariant four-dimensional Bethe-Salpeter
equations (BSE) \cite{BETSAP1951}. In most studies the kernel of the
BSE is approximated by a single boson exchange
(ladder-approximation). Besides, all single particle propagators are
very often replaced by the free ones. In the thesis we move beyond
bare ladder approximation and include the fully dress single particle
GFs.

We  considered simple super-renormalizable model of three massive
fields with cubic interaction $\phi_i^2\phi_3\,;i=1,2$ (the massive
Wick-Cutkosky model). Having the propagators calculated within  the
bare vertex truncation of the SDEs we combine them with the dressed
ladder BSE for the scalar s-wave bound state amplitudes, following
the treatment of ref.~\cite{SAUADA2003} where the spectral technique
was used to obtain the accurate results directly in Minkowski space.
In this paper the analytic formula has been derived for the kernel of
the resulting equation which significantly simplifies the numerical
treatment. For the extensive, but certainly not exhaustive,
up-to-date review of methods to treat the scalar models see also
\cite{SAUADA2003}.

Analogously to the treatment of the SDEs reviewed above, the BSE
written in momentum space is converted into a real integral equation
for a real weight function. This then allows us to treat the ladder
BSE in which all propagators (of constituents and of the exchanged
particle) are fully dressed. Having solved equations for BSE spectral
functions, one can easily determine the BS amplitudes in an arbitrary
reference frame.

The BS amplitude for bound state $(\phi_1,\phi_2)$ in momentum space is
defined through the Fourier transform of
\be
\langle 0|T\phi_1(x_1)\phi_2(x_2)|P\rangle = e^{-iP\cdot X}
\int{d^4p\over (2\pi)^4} e^{-ip\cdot x} \Phi(p,P)\ .
\ee
The corresponding scalar BS vertex function
$\Gamma=iG_{(1)}^{-1}G_{(2)}^{-1} \Phi $ satisfies
\be
\label{bse}
\Gamma(p,P) = i\int{d^4k\over(2\pi)^4}\, V(p,k;P)
G_{(1)}(k+P/2)G_{(2)}(-k+P/2) \Gamma(k,P)\, ,
\ee
where $V$ is the kernel of the BSE and bracketed subscript label the
particles.
%
When the mass of exchanged particle is nonzero,  the appropriate
integral representation for the BS vertex function is two
dimensional:
\be
\label{nula}
\Gamma(p,P)= \int\limits_{-1}^1 dz \,
\int\limits_{\alpha_{min}(z)}^{\infty}d\alpha\, \,
  \frac{\rho^{[n]}(\alpha,z)}{[\alpha-(p^2+zp\cdot P+P^2/4)-i\epsilon]^n} \, \, .
\ee
The positive integer $n$ is a dummy parameter constrained only by
convergence of the BSE, $n=2$ was found to be the most suitable
choice. The BSE can be  converted into the following real integral
equation for the real spectral function:
\be
\label{majn} \rho^{[2]}(\alpha',z')=\frac{g^2}{(4\pi)^2}
\int\limits_{-1}^{1}dz \int\limits_{\alpha_{min}(z)}^{\infty}
d\alpha\, \, V^{[2]}(\alpha',z';\alpha,z)\, \rho^{[2]}(\alpha,z)  \,
.
\ee
The BSE interaction kernel is $V(p,k;P)=g^2G_{(3)}(p-k)$ in the
dressed ladder approximation, in which  all propagators are fully
dressed. For this case the kernel $V^{[2]}(\alpha',z';\alpha,z)$ in
spectral equation (\ref{majn}) has been derived in \cite{SAUADA2003}.
The explicit  formula  for $V^{[2]}$ involves the integrals over the
spectral  function $\bar{\sigma}_{(i)}$ from the SR of the propagator
of the i-th field.

In the ladder approximation the obtained energy spectrum  agree with
ones obtained by other techniques
\cite{KUSIWI1997,NIETJO1996,AHLALK1999,KUWI1995}. In the dressed
ladder approximation there is no reliable published result  to
compare with. However, the qualitative agreement with paper
\cite{AHLALK1999} was found: The critical value of coupling constant
$g_{crit}$ (defined by  the value of $g$ for which the
renormalization constant becomes zero: $Z(g_{crit})=0$) determined
from SDEs gives the domain of applicability of the BSE. The couplings
below the critical one allow only solutions for relatively weakly
bound states.

In the original Wick-Cutkosky model \cite{WICK}  the exchanged boson
is massless and no radiative corrections are considered. This model
is particularly interesting because it is the only example of the
nontrivial BSE which is solvable exactly \cite{WICK}. For this model
the s-wave bound-state PTIR reduces to one variable spectral integral
and the equation for the spectral function simplifies. The resulting
weight functions $\rho(z)$ for various  are displayed in
Fig.~\ref{fweighty} for several values of $\eta$.
\begin{figure}[t]
\centerline{  \mbox{\psfig{figure=WCrho.eps,
height=7.0truecm,width=8.0truecm,angle=0}}
\mbox{\psfig{figure=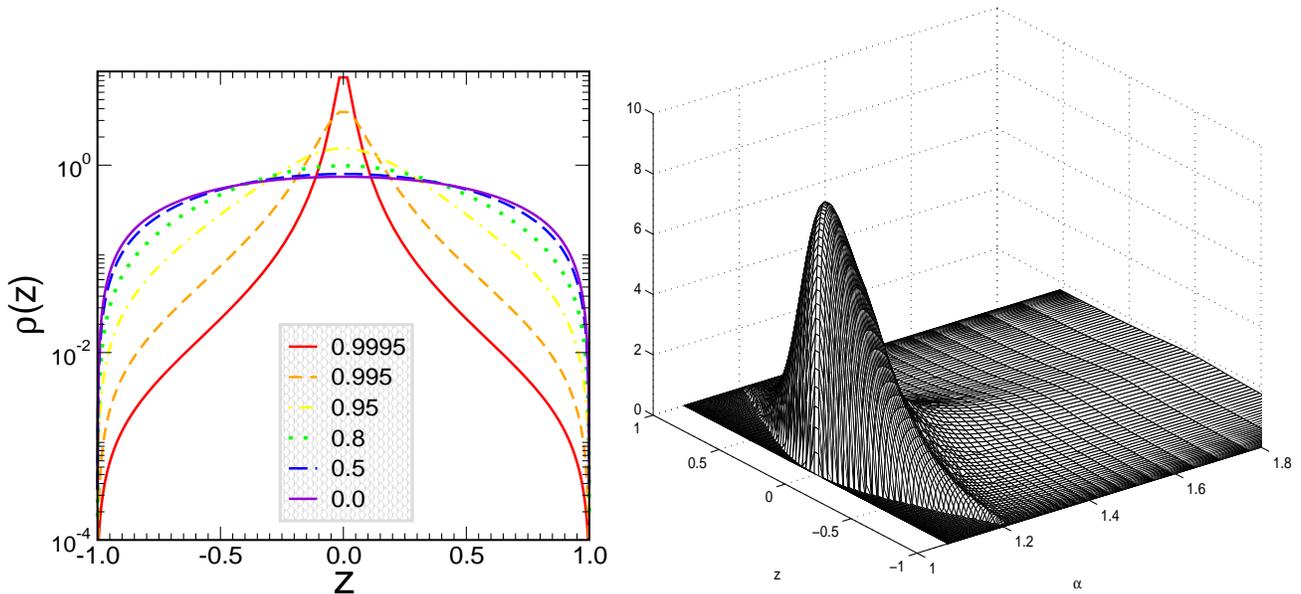,
height=8truecm,width=9truecm,angle=0}}}
\caption[99]{\label{fweighty}
Left figure displays the weight function $\rho(z)$ for the original WCM
$(m_3=0)$ for several values of $\eta=\sqrt{P^2}/2m$. The right figure
displays rescaled weight function for the massive WCM
for $m_3=0.1m$ and $\eta=0.95$. }
\end{figure}

The electromagnetic form factors parametrize the response of bound
systems to external electromagnetic field. The calculation of these
observables within the BS framework proceeds along the Mandelstam's
formalism \cite{MAN1955}.

The current conservation implies the parametrization of the current
matrix element $G^{\mu}$ in terms of the single real form factor
$G(Q^2)$
\be  \label{formf}
G^{\mu}(P_f,P_i)=G(Q^2)(P_i+P_f)^{\mu} \, ,
\ee
where $Q^2=-q^2$, so that  $Q^2$ is positive for  elastic
kinematics.

\begin{figure}[t]
\centerline{  \mbox{\psfig{figure=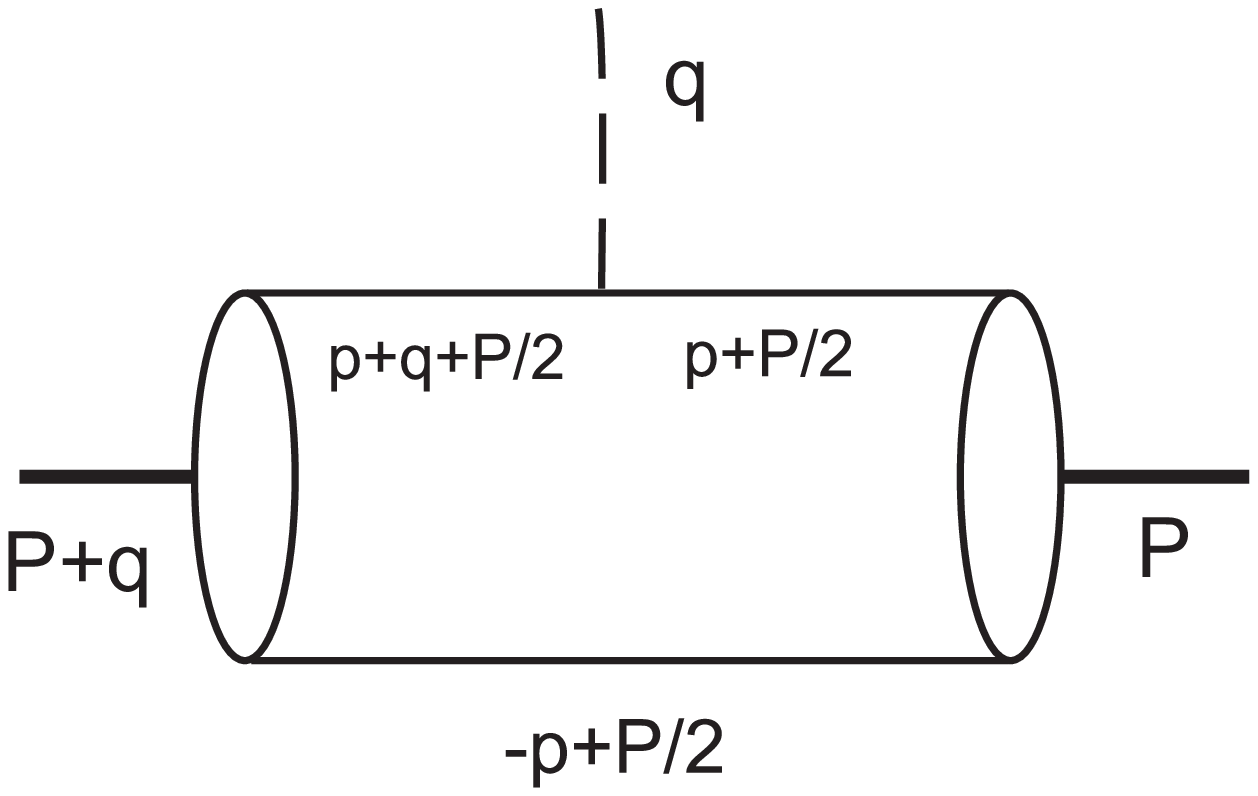,height=4.0truecm,angle=0}} }
\caption[99]{\label{figGmu} Diagrammatic representation of the electromagnetic  current bound state
matrix element.}
\end{figure}

The matrix element of the current in relativistic impulse
approximation (RIA) is diagrammatically depicted in
Fig.~\ref{figGmu}. The matrix element is  given in terms of the BS
vertex functions as
\bea
&&G^{\mu}(P+q,P)=i\, \int\frac{d^4p}{(2\pi)^4}\,
\bar\Gamma(p+\frac{q}{2},P+q) \nonumber\\
&& \left[ D(p_f;m_1^2) j_1^{\mu}(p_f,p_i) D(p_i;m_1^2) \,
D(-p+P/2;m_2^2) \right] \Gamma(p,P) \, ,
\label{chargeff}
\eea
where we denote $P=P_i$ and $j_1^{\mu}$ represents one-body current
for particle $\phi_1$, which for the bare particle reads
$j_1^{\mu}(p_f,p_i)= p_f^{\mu}+p_i^{\mu}$, where $p_i,p_f$ is initial
and final momentum of charged particle inside the loop in
Fig.~\ref{figGmu}.

We have rewritten  the r.h.s.\ of Eq.~(\ref{chargeff}) directly in
terms of the spectral weights of the bound state vertex function. It
allows the evaluation of the form factor by calculating the
dispersion relation:
\be \label{drforf}
G(Q^2)=\int d \omega \frac{\rho_G(\omega)}{Q^2+\omega-i\epsilon}
\ee
without having to reconstruct the vertex functions $\Gamma(p,P)$ from
their spectral representation.

\section{Gauge theories}

Large part of the thesis deals with gauge theories, with main stress
on the strong-coupling Quantum Electrodynamics, treated in Chapter 6.

In the last Chapter of the thesis we employ the generalized
spectral decomposition for the gluon propagator (formally
identical to the Lehmann one, but with the spectral function which
is not required to be positive) to $\sigma(\omega)$ from recent
Euclidean lattice and SDE results and analytically continue them
into the timelike region.

\subsection{SDEs of one-flavor strong QED}

\begin{figure}[t]
\centerline{\epsfig{figure=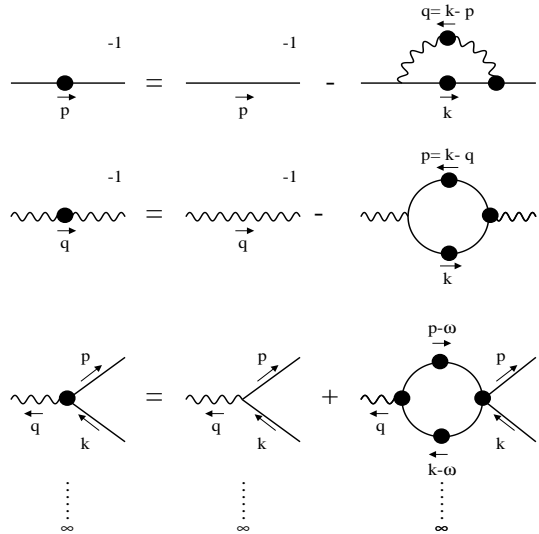,width=7truecm,height=7truecm,angle=0}}
\caption[caption]{Diagrammatical representation of the
Schwinger-Dyson equations in the quantum electrodynamics. The black
blobs represent the full propagators and vertices. Wavy lines stand
for the photon and  full lines for the fermion
propagator.\label{setSDE}}
\end{figure}

The brief textbook derivation of the  Schwinger-Dyson equations
for QED (Fig.~\ref{setSDE}) is given already in the introductory
Chapter 2 of the thesis.  In Chapter 6 we investigate SDEs of the
3+1 dimensional QED, in a first attempt to move beyond the class
of scalar models. We review the methods and results of papers
\cite{SAULIJHEP,SAULIRUN}, in which the solutions for this theory
in Euclidean and Minkowski space were compared for the first time.
Since no non-trivial non-perturbative effects (i.e., effects not
known from PT) were found in the photon propagator, our attention
is mainly concentrated on the SDE for the fermion one. We  deal
mostly with the strong coupling regime, which is far from the
``real-life QED'', for which  $\alpha_{QED}$ is small (in
experimentally accessible energy region)  and use of the PT is
fully justified. Of course, all our solutions agree in weak
coupling limit with the perturbation theory (PT).

The strong-coupling Abelian dynamics was considered as one of the
candidates for explanation of the electroweak symmetry breaking
\cite{PAGELS}. Although we do not think this to be a realistic
model for a mass generation of the SM fermions, it could be a
candidate for the strong coupling sector of theories like
Technicolors, e.g., Slowly Walking Technicolors
\cite{WEI1979,HOL1985,CHIV2001,APP2003}, for which the LA of a
QED-like theory seems to provide a  reasonable model of
(Techni)lepton propagators. Moreover, the strong-coupling Abelian
model is not only a suitable playground for studies of
supercritical phenomena like dynamical mass generation, it is
suitable also for investigations of the analytical structure of
the fermion propagator \cite{SAULIJHEP,FUKKUG1976}. It was 30
years ago when Fukuda and Kugo \cite{FUKKUG1976} observed the
disappearance of the real pole of the fermion propagator in the
ladder QED and it was argued that this is a signal for confinement
of the fermion. Again, the main motivation is the simplicity of
the model, in alternative more complicated models the
non-perturbative phenomena are even more difficult to study
quantitatively.

3+1 QED has been studied frequently, mostly in Euclidean formalism
employing various approximations. The truncation of the SDEs should
not violate the gauge invariance. At least two technically different
ways were proposed to deal with SDEs of QED. Ball and Chiu
\cite{BALCHI1980} derived the formula for the proper vertex which--
being written in the terms of propagator functions $A$ and $B$--
closes the SDEs system in a way respecting WTI. The second approach
is known as the ``Gauge Technique''. In this framework the spectral
Anzatz is made for the untruncated vertex function, i.e., for
$G^{\mu}=S\Gamma^{\mu}S$. The technical advantage of the Gauge
Technique is that the resultant equation for the fermion propagator
is linearized in a spectral function. On the other hand, this is one
of essential weaknesses of this approach: it cannot be true in
general, even for the case of electron propagator. It is clear that
beyond the simplest approximations employed in the literature
\cite{SALDEL1964,DELWES1977,WEST2,DELZHA1984,HOS2002} such
linearization does not take place.

The first part  of Chapter 6 of the thesis is devoted to the solution
for the electron propagator in the {\it ladder approximation} (LA),
in the second part extension to the {\it unquenched} (with the photon
polarization included) case is made. In the latter case the running
coupling is considered {\it self-consistently}: we follow the paper
\cite{SAULIRUN}, where the fermion mass and photon polarization
function have been calculated by solving the coupled SDEs for
electron and photon propagators in the Landau gauge.

In both approximations we are looking both for solutions with zero
and non-zero bare electron mass in the Lagrangian. In the first case
the chirally symmetric solution (with massless electrons) always
exists. In the Euclidean formalism we obtain, in agreement with many
previous studies, for sufficiently large $\alpha$ also non-trivial
solution for the mass function $M$. On the other hand, no such
solution was found in our spectral Minkowski approach. This is a
strong indication that the  fermion mass function has in this case a
complicated analytical structure, which is not reflected by
assumptions of our spectral Anzatz.

In explicit chiral symmetry breaking (E$\chi$SB) case, in which the
non-zero electron mass exists from the very beginning,  both
approaches-- Minkowski and Euclidean-- offer approximately the same
results (in the spacelike domain, of course) in the regime where
both solutions were obtained. It is interesting that in the
unquenched approximation the E$\chi$SB spectral solution (with
explicit mass term) fails almost exactly for the same value of the
coupling at which the S$\chi$SB solution occur.

\begin{figure}
\vspace*{-2.0truecm}
\begin{center}
\mbox{\epsfig{figure=m400.eps,height=6.0truecm,width=6.0truecm,angle=0}}
\caption[99]{\label{fweighty} Figure presents the Minkowski
solutions of the SDE for timelike momenta, i.e., in the range
where solutions are not available  in much more common Euclidean
approach. The electron mass function $M$ [from the propagator $S=
F(p^2)/(\slashed{p}-M(p^2)]$  calculated in the ladder
approximation in the Landau gauge is shown. The renormalization is
chosen such that $M(-10^8)=400$ and the lines are labelled by the
coupling constant $\alpha$. The cusps correspond to the physical
pole masses, while the absorptive parts of $M$ correspond to the
smooth lines. The results in the spacelike regime agree with the
ones obtained in Euclidean calculations.}
%
%
\mbox{\epsfig{figure=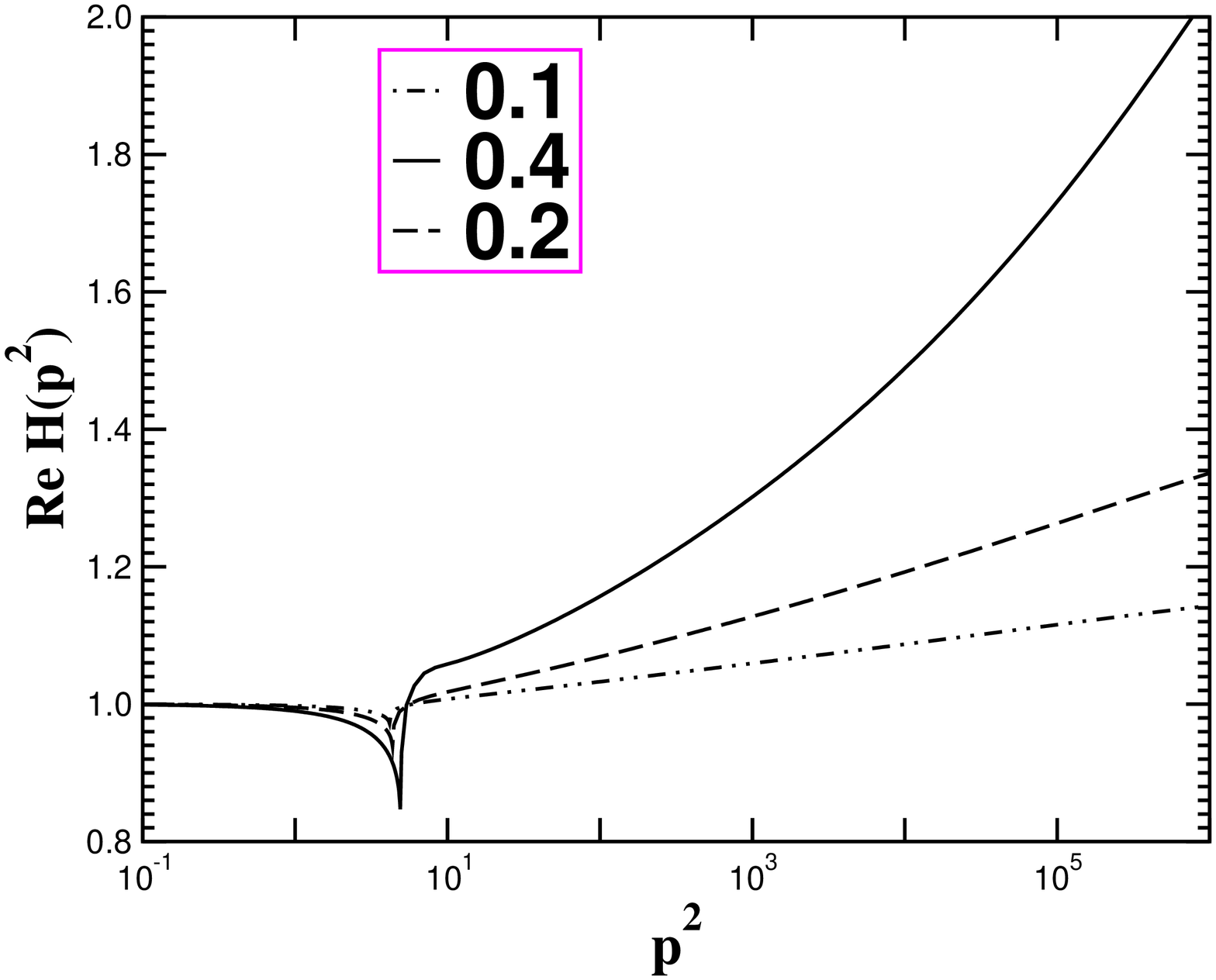,
height=7truecm,width=7truecm,angle=270}}
\caption[99]{Figure shows timelike behavior of the charge
renormalization function, i.e., the running coupling constant
normalized to 1 at zero momentum, as it was obtained from the SDEs
for the unquenched QED in the Landau gauge. The vertex function
was approximated by the bare vertex  $\gamma_{\mu}$. The
down-oriented peak corresponds to the threshold $4m^2$, the mass
of the fermion was renormalized such that $M(0)=1$. The lines are
labelled by the  renormalized couplings at zero momenta:
$\alpha(0)=0.1, 0.2, 0.4$.}
\end{center}
\end{figure}

In order to compare the Minkowski solution with some recent Euclidean
results \cite{HSW1995,HSW1997,KONNAK1992} the same approximations has
to be made and the same schemes has to be picked up. By these we mean
the truncation of the SDEs, the same choice of the gauge and  of the
renormalization scheme. The resulting integral equations are solved
in their full form, the linearization inherent to the Gauge technique
is avoided. The unitary equations and dispersion relations for
self-energies are derived both for ladder and unquenched
approximations. For both of them also the Euclidean formalism is
reviewed. The numerical results are briefly summarized below.

The solutions of the SDEs were obtained for several values of the
coupling constant and for several renormalization choices. In the
ladder approximation the expected damping of the  mass function to
its negative values is observed in  the supercritical phase of QED
\cite{HSW1997,FUKKUG1976}.  The Euclidean and Minkowski solutions for
the fermion mass functions are compared in several figures. The
confirmation of disappearance of the real pole (first indicated in
\cite{FUKKUG1976}) in the supercritical ladder QED is  one of the
main physical results of this section and of paper \cite{SAULIJHEP}.
We also argue that the fermion SR is absent in the supercritical QED.
We have but a numerical evidence for such a statement, hence it
should remit to further investigation.

When the vacuum polarization  effects are  taken into account
self-con\-sis\-tent\-ly we confirm triviality of QED. It implies that
some kind of regularization is needed beyond and irrespective of the
renormalization technique employed. We explain how to introduce such
cut-off in our Minkowski spectral treatment. In numerics, this
cut-off is taken to be $\Lambda^2=10^7M^2(0)$, where $M(0)$ is the
infrared electron mass. The numerical solution fails when $\alpha$
exceeds the value $\alpha_R(0)=0.41$. Note that at ultraviolet region
the running coupling $\alpha_R(\Lambda^2)$ is approximately 2.6 times
larger then $\alpha_R(0)$. We are not able to see Landau pole
directly (it corresponds to $\alpha_{\mp}\rightarrow \pm \infty$).
However, the observed large growth of the running coupling and of its
derivative can be understood as the evidence for the Landau
singularity somewhere above the cut-off.

\section{QCD}

The non-Abelian character of the QCD makes it difficult to convert
the momentum SDEs into equations for spectral functions. We were not
able to do this yet (the main obstacle is the ghost SR, mainly due to
zero momentum behavior). Therefore, we first briefly review the
symmetry preserving {\it gauge invariant} solution obtained by
Cornwall two decades ago \cite{COR1982}. To our knowledge this is the
best published example, in which the behavior of the QCD Green
function in the whole range of Minkowski formalism is addressed
within the framework of the SDEs. Instead of solving the SDEs we use
the generalized spectral representation to fit the spectral function
to Euclidean solutions obtained in recent lattice simulations and in
the SDEs formalism.

The Quantum Chromodynamics (QCD) is the only experimentally studied
strongly interacting relativistic quantum field theory. This
non-Abelian gauge theory with a gauge group $SU(3)$ has many
interesting properties. The dynamical spontaneous breaking of chiral
symmetry explains why the pions are light, identifying them with the
pseudo-Goldstone bosons associated with the symmetry breaking of the
group $SU(2)_L\times SU(2)_R$ to $SU(2)_V$ (in flavor space).
Asymptotic freedom \cite{GROWIL1973,POL1973} implies that the
coupling constant of the strong interaction decreases in the
ultraviolet region. For less than 33/2 quark flavors the QCD at high
energy becomes predictable by the PT. However, in the infrared region
the PT does not work and non-perturbative techniques have to be
applied.

One of the most straightforward non-perturbative approaches is a
solution of the SDEs for QCD. The extensive studies were
undertaken for a quark SDE, based on various model assumptions for
a gluon propagator. These approximate solutions, often accompanied
by a solution of the fermion-antifermion BSE for meson states,
have become an efficient tool for studies of many non-perturbative
problems, e.g., the  chiral symmetry breaking, low energy
electroweak hadron form factors, strong form factors of exclusive
processes, etc (see reviews
\cite{ROBERTS,ALKSMEK2001,MARROB2003,ROBSMI2000} and also recent
papers \cite{FISALK2003,BENDER,BICEST2003,BIC2004,FISALKCON}).

However, to take gluons into account consistently  is much more
difficult than to solve the quark SDE alone. The SDE for the gluon
propagator   is more non-linear than the quark one. Moreover, the
Faddeev-Popov ghosts have to be included \cite{FADPOP1967} in a class
of Lorentz gauges. In recent papers
\cite{BLOCH2003,FISALK2003,ATKBLO1998,KONDO2003,AGUNAT2004} studies
of the coupled SDEs for gluon and ghost propagators in  the Landau
gauge in Euclidean space were performed in various approximations.

In Chapter 8 we  consider the implication of analyticity for the
solution of SDEs for gluon propagator. As in previous Chapters we
assume analyticity of the propagators. The generic SR for the
renormalized gluon propagator in the Landau gauge reads
\bea G^{\alpha,\beta}_{AB}(q)&=&\delta_{AB}\,
\left[-g^{\alpha\beta}+\frac{q^{\alpha}q^{\beta}}{q^2}\right]\,
G(q^2) \, , \nn\\
G(q^2)&=&\int\limits_{0}^{\infty} d\omega\,
\frac{\sigma(\omega;g(\xi),\xi)}{q^2-\omega+i\epsilon} \, ,
\label{gluesp} \eea
where only here  full dependencies of the continuous function
$\sigma(\omega;g(\xi),\xi)$ are indicated explicitly.  The Ansatz
(\ref{gluesp}) should be consider the generalized spectral
representation, since we do not assume (and do not obtain) the
spectral function $\sigma(\omega)$ positive for all values of
$\omega$.

\begin{figure}
\centerline{
\epsfig{figure=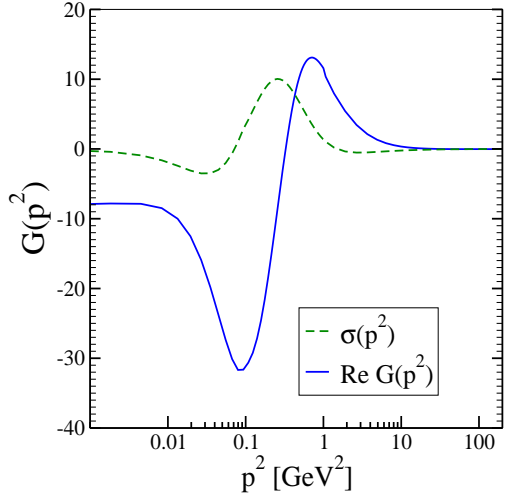,width=8truecm,height=10truecm,angle=270}}
\caption[caption]{ \label{gluontime}The gluon propagator for
timelike momenta.  }
\end{figure}

To find $\sigma(\omega)$ we first represented some recent lattice
\cite{BOBOLEWIZA2001} and   SDEs results \cite{FISALK2003} (both
defined in spacelike region) by an analytical formula. Our spectral
decomposition is fitted to these ``data'' for spacelike momenta and
then predicts the gluon propagator in the timelike region. The
solutions are plotted in Fig.~\ref{gluontime}. The spectral function
has a smooth peak around  $p^2=(0.7{\rm GeV})^2$ with the width
$\approx 1$GeV.  It becomes negative for asymptotically large $p^2$,
as  expected already  from the PT. The gluon propagator should
describe the confined particle, so the ``unusual'' shape (violation
of  the spectral function positivity) is in accord with our physical
expectations.

\section{Yukawa theory}

In Chapter 7 of the thesis the SDEs of the massive Yukawa theory are
studied. The Yukawa interaction appears in  many models of Nature.
Original motivation for developing the Yukawa theory (YT) was
modeling of the nuclear forces.  Several decades from its birth the
Yukawa interaction has become the firm part of the Standard Model and
of many models beyond it. In alternatives to the minimal SM, such as
its SUSY extensions \cite{WEISUSY}, the Little Higgs models
\cite{ACKN2002}, \cite{LOGAN}, the Zee model of the neutrino sector
\cite{ZEE1980}, \cite{CHANZEE1999}, and in many others, the
interaction of spinless particles with matter generates the masses of
particles. This can happen at tree level due to the nonzero vacuum
expectation values of the Higgs fields (like in the SM) or the masses
can be generated dynamically through the radiative corrections. Of
course, the combination of both effects is also possible.

As for the vectorial interaction from the previous Chapter, this
study of the YT SDEs should be regarded as a first step in
building intrinsically non-perturbative methodological tools
within the spectral formalism in Minkowski space. Such tools
should next help to reveal the quantitative face of dynamical
phenomena, such as the triviality of this QFT, the quantum
stability or the dynamical mass generation for the YT with a given
parameter space. Some attempts to study such phenomena in the YT
(or the SUSY YT) have already been made
\cite{SAULI2004,BASLOR1999}.

The SDEs for the YM are derived from the following Langrangian:
\bea
\label{yukawamodel}
{\cal L}&=&i\bar{\Psi}_0\slashed{\partial}\Psi_0 -
m_{0}\bar{\Psi}_0\Psi_0 + \frac{1}{2}\partial_{\mu}\Phi_0\,
\partial^{\mu}\Phi_0 -\frac{m_{0\phi}^2}{2}\Phi_0^2
\nn \\
&-&ig_{0}\, \bar{\Psi}_0\gamma_{5}\Psi_0 \Phi- h_0\Phi_0^4 \,
,
\eea
where we assume (mainly for historical reasons) that the field
$\Phi_0$ is pseudoscalar. If the Yukawa coupling  $g_0$ is nonzero,
the term $h_0\Phi_0^4 $ has to be present in the Lagrangian due to
the general requirement of renormalizability. For the sake of
simplicity, the Yukawa vertex is modeled by its tree approximation
and  it is assumed that the renormalized constants satisfy $h<<g$,
therefore we also neglect the quartic meson self-interaction. For the
renormalization the momentum subtraction scheme is  used. The
numerical results are given for the absorptive and dispersive parts
of the self-energies. The SDEs results are in agreement with our
expectation supported by the perturbation analysis: one gets rather
small corrections to the fermion self-energy, whereas the boson
self-energy exhibits a quadratic behavior for large momenta squared
(quadratic divergences turn quadratic dependence after the
renormalization).


\section{Summary, conclusion and outlook}

A connection between the Euclidean space and the physical
Minkowski space is an important question, in field theory in
general and in the SDE approach in particular. We have explicitly
demonstrated (albeit on simple models and employing some
simplifying approximations) that SDEs studies  are feasible
directly in Minkowski space. A reasonable agreement of the
spectral SDEs solutions (at spacelike momenta) with the ones
obtained by more conventional strategy in Euclidean space gives us
strong belief in a relevance of developed methods.

We started an introduction of the spectral concept into the
formalism of SDEs with specific attention to scalar models and
QED. We performed studies of the strong-coupling QED in various
approximations. One of them is unquenched QED, in which the
running of the coupling constant has been correctly taken into
account. In this case the QED triviality plays its crucial role
for asymptotically large spacelike momenta. We explain how to deal
with this trivial theory within the formalism of spectral
representations and dispersion relations.

The strong coupling QED is often regarded as an ideal pedagogical
tool  for SDE studies and their application to QCD. We also attempted
to extend the discussion to QCD, for which the direct Minkowski space
formulation and solution is not yet fully developed, since the theory
is much more non-linear and complicated.

While many important steps have been made, it is obvious that much
more needs to be done. Up to now all studies of profoundly
non-perturbative phenomena, such as the chiral symmetry breaking or
dynamical mass generation,  were carried out in Euclidean space. It
is not quite clear how to perform such symmetry breaking SDE studies
directly in Minkowski space.

Another interesting perspective is to further explore the timelike
infrared behavior of Green functions obtained by analytical
continuation of lattice results. We offered a solution for gluon
propagator in the Landau gauge, but information on the timelike
structure of the quark propagator is currently unavailable.

We also solved the Bethe-Salpeter bound-state equation in (3+1)
Minkowski space within the spectral framework in the dressed
ladder approximation. It would be interesting to extend this
technique to more complicated BS kernels: e.g., to include
cross-box contributions, $s$ and $u$ channel interactions etc.

Further developed, the Minkowski space spectral technique should
become rather efficient tool of hadronic physics. It was already
demonstrated that obstacles due to the fermionic degrees of
freedom can be overcome. The solution of the spinless bound state
of two quarks interacting via dressed gluon exchange (a pion) is
under auspicious consideration.

\newpage

\begin{center}{\bf Author's papers}:\end{center}

\end{document}